# Is PMBOK Guide the Right Fit for AI? Re-evaluating Project Management in the Face of Artificial Intelligence Projects

Alexey Burdakov
Intel Corporation
Munich, Germany
alexey.burdakov@intel.com

Max Jaihyun Ahn
Intel Corporation
Seoul, South Korea
max.jaihyun.ahn@intel.com

*Abstract*—This paper critically evaluates the applicability of the Project Management Body of Knowledge (PMBOK) Guide framework to Artificial Intelligence (AI) software projects, highlighting key limitations and proposing tailored adaptations. Unlike traditional projects, AI initiatives rely heavily on complex data, iterative experimentation, and specialized expertise while navigating significant ethical considerations. Our analysis identifies gaps in the PMBOK Guide, including its limited focus on data management, insufficient support for iterative development, and lack of guidance on ethical and multidisciplinary challenges. To address these deficiencies, we recommend integrating data lifecycle management, adopting iterative and AI project management frameworks, and embedding ethical considerations within project planning and execution. Additionally, we explore alternative approaches that better align with AI's dynamic and exploratory nature. We aim to enhance project management practices for AI software projects by bridging these gaps.

*Keywords—PMBOK Guide, artificial intelligence, software development, project management*

## I. Introduction

The artificial intelligence (AI) software market is rapidly expanding, reflecting the increasing integration of AI solutions in diverse sectors. The potential of AI drives this growth [1] to enhance efficiency, optimize decision-making, and foster innovation. Industries ranging from healthcare to finance and manufacturing are leveraging AI to gain competitive advantages, leading to an unprecedented surge in the demand for AI software projects. Recent studies [2] [3] show that the global AI software market has grown exponentially in recent years and continues its trajectory as organizations increasingly adopt AI-driven solutions.

An AI software project is an undertaking that aims to deliver a software product or service that embeds AI functionality for use by humans or machines. This project type differs significantly from traditional software projects and projects in general [4]. The differences stem from several distinct features paramount to AI software development.

The Project Management Body of Knowledge (PMBOK) Guide [5] is a widely recognized framework that provides principles and guidance for managing projects effectively. While the PMBOK Guide offers robust methodologies for general project management, it does not necessarily address the unique needs of AI software projects.

The related work in this area studies the difference between Traditional IT projects and AI projects [4]. This study underscores the need for tailored methodologies to better address the demands of AI initiatives. The article [6] emphasizes the unique management challenges in AI projects, including the need for interdisciplinary collaboration, iterative development processes, and addressing ethical considerations. The work [7] highlights some key challenges of Data Science (DS)/AI projects and offers elements of a framework to address them. The case study [8] provides four strategies to help organizations manage their AI projects better. The work [9] explores and investigates project management methodology for artificial intelligence (AI) transformation projects. The Systematic literature review in [10] analyzes 54 AI project management challenges and state-of-the-art project management frameworks. However, the existing work does not consider the PMBOK Guide framework, which could be a foundational approach if tailored to AI projects' specific needs. While the PMBOK Guide is widely used, its applicability to AI projects remains underexplored, as shown by the analysis of the current literature.

Practitioners in AI software development stand to benefit from the PMBOK Guide tailoring recommendations that specifically address their projects' unique challenges and requirements. This paper seeks to fill this gap by studying the distinct needs of AI software projects and providing recommended guidance to better align project management practices with these needs.

The authors combined case studies and a literature review to explore AI software project (AISP) management. The case studies of open source code projects Intel® Geti™ [11] [12] [13], OpenVINO® [14], Training Extensions [15], Datumaro [16], Anomalib [17] [18], Explainable AI Toolkit [19] and other AI projects provide real-world insights. The case studies were selected based on their relevance, diversity of applications, and real-world implementations. At the same time, the literature review synthesizes existing research on methodologies and trends. The criteria for literature inclusion were based on relevancy to one or more of the following criteria: AI project's distinct features, AI software projects, and application of the PMBOK Guide. The search was done through publication databases, search engines, and Generative AI helpers.

The paper is structured as follows: Section 2 explores the unique features of AI projects and how they differ from traditional software projects. Section 3 overviews the PMBOK Guide framework and its core principles. Section 4 evaluates how well the PMBOK Guide aligns with the unique features of AI software projects, highlighting gaps and providing recommendations for advancing project management practices to better support AI software development. Finally, Section 5 presents conclusions.

## II. ARTIFICIAL INTELLIGENCE (AI) PROJECT FEATURES

Based on our experience with over 10 AI projects and a comprehensive literature review, we hypothesized that the key features of AI projects are data dependency [6] [20] [21], iterative development [22] [23] [8] [24] [25], uncertainty [26] [27] [28] and experimentation [5] [29] [30], the need for specialized expertise [4] [31], and ethical considerations [32] [33] [34] [35] [29]. Below, we formulate the description of these project features (PF):

- Data Dependency (PF1): AI projects rely heavily on high-quality, relevant datasets for training and performance. Ensuring data quality, accuracy, completeness, and bias mitigation is critical. Data preparation is time-consuming, and handling sensitive data requires robust security measures and legal compliance.
- Uncertainty and Experimentation (PF2): AI model outcomes are unpredictable, necessitating trial and error with algorithms, hyperparameters, and datasets. This process requires tolerance for failure and adequate time for addressing issues ("black box problem"), retraining models, and conducting experiments. Defining requirements and timelines can be challenging due to the lack of direct competitors or precise specifications.
- Iterative Development (PF3): AI development demands an experimental approach with unpredictable timelines, challenging traditional sprint planning. Model tuning and experimentation often require mid-course adjustments or span multiple sprints, necessitating flexibility and continuous adaptation. This differentiates AI software projects from traditional ones (e.g., user interface development), which align well with Agile.
- Specialized Expertise (PF4): AI projects require multidisciplinary teams with specialized skills in data preprocessing, algorithm design, and deployment. Effective communication between technical and non-technical members (including subject matter experts) is crucial. Advanced tools and technologies, such as AI/ML frameworks, big data platforms, and cloud services, are essential. The rapidly evolving AI landscape is vital for talent acquisition and continuous upskilling.
- Ethical Considerations (PF5): Ethics play a pivotal role in AI development [32], forcing projects to avoid discrimination, ensure fairness, and maintain transparency. Privacy and data protection are fundamental, requiring robust safeguards. The design of AI systems shall consider safety and robustness, with human oversight and accountability to align decisions with human values and judgment.

To validate this hypothesis, we surveyed experts with significant AI project experience, asking them to rank the importance of each feature PF1-PF5 on a scale from 1 to 10 and suggest additional features. Analysis of 39 data points for 6 AI software projects (AISP1-AISP6), as shown in Figure 1.

It indicates that experts strongly support our hypothesis, particularly emphasizing the importance of data dependency, uncertainty and experimentation, and the need for specialized expertise. The PF3 iterative approach was ranked comparatively lower as Agile was still applicable by some respondents if appropriately applied to the lifecycle.

## III. WHAT IS PMBOK GUIDE?

PMBOK® (Project Management Body of Knowledge) Guide is a globally recognized standard for project management. It originated in 1981 and has evolved through several editions, the latest being the 7th edition [5] published in 2021.

This guide provides a comprehensive framework of principles, processes, and best practices for effectively managing projects. It has gained significant popularity due to its structured approach, global recognition, and role in professional certifications like the Project Management Professional (PMP).

Earlier, many publications criticized the PMBOK Guide framework for being too rigid, especially compared to agile methodologies [36], which emphasized flexibility, iterative processes, and rapid delivery, widely adopted in software development. However, these concerns were addressed mainly by adopting a more flexible and adaptable approach, recognizing the growing importance of agile and hybrid methodologies in today's dynamic business environment. The 7th edition focuses on principles and outcomes rather than prescriptive processes, allowing project managers to incorporate a variety of methodologies—agile, traditional, or hybrid—to meet the needs of their projects best and deliver value to stakeholders.

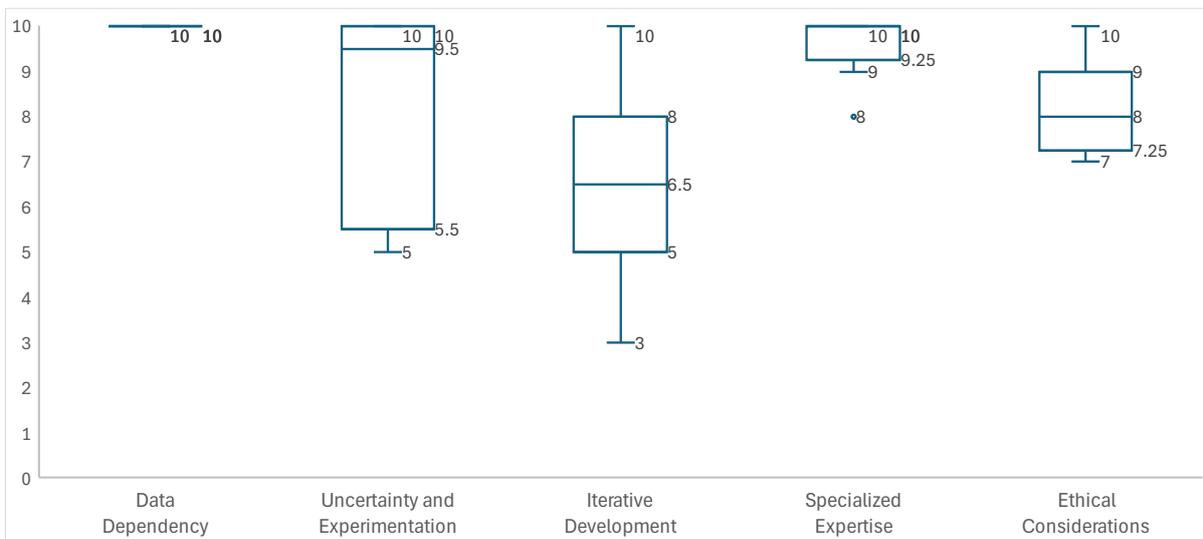

Figure 1 Expert Rankings of Key Features in AI Projects

The PMBOK Guide introduces eight performance domains [5], which represent key focus areas for delivering successful project outcomes. A performance domain is a group of activities critical to achieving project success, emphasizing the holistic integration of project components rather than rigid processes. These adaptable domains allow project teams to address each environment's dynamic and unique needs. Additionally, it provides tailoring recommendations (Section 3.5 in [5]), encouraging practitioners to adapt these tools and practices to suit their projects' specific contexts and requirements. This flexible approach ensures relevance across diverse industries and project types.

IV. How well does the PMBOK Guide fit the AI project features?

While the PMBOK Guide [5] offers a comprehensive framework for project management and outlines principles applicable to a wide range of projects, it does exhibit some notable gaps. To address these gaps, we have examined the AI software project features, specifically PF1-PF5 (refer to Section 2), and conducted a detailed analysis across the PMBOK Guide performance domains. Sections 4.1-4.5 present an analysis of each feature (PF1-PF5) and provide tailored recommendations to bridge the identified gaps.

A. *Data Dependency (PF1)*

*1) Analysis*

Data is a critical component of AI projects and significantly influences the success of development efforts. A dataset is a collection of structured or unstructured data used to train or evaluate an AI model's performance. The datasets can include various data types, such as images, alphanumeric text, audio, telemetry, and video, and may or may not contain personal data.

The quality, quantity, and relevance of data directly impact the performance of AI models. High-quality data enables more accurate predictions and insights, while poor-quality data can lead to suboptimal or even erroneous outcomes [4] [8].

Data-driven projects, particularly those involving AI models, shall include the following considerations [6] [20] [21]:

- Central Role of Data: Focus on managing the data lifecycle, including collection, cleaning, and preparation, ensuring alignment with project objectives, and applying tools or best practices for data-driven workflows.
- Data Quality: Ensuring accuracy, completeness, and bias mitigation by using bias detection and mitigation techniques, evaluating dataset relevance and data balance, and preventing errors in training datasets [25].
- Data Ownership and Licensing: This involves identifying ownership rights, navigating licensing agreements for datasets, and ensuring compliance with intellectual property rights like copyrights and patents.
- Data Privacy and Compliance: Adherence to regulations such as GDPR or CCPA, using anonymization and pseudonymization techniques, and securing necessary consent for data collection and processing.
- Liability and Accountability: Define responsibility for data-related issues, mitigate risks like data breaches or copyright violations, and address risks from data quality affecting AI performance.
- Dataset Procurement: Establishing processes for sourcing datasets, including evaluating data providers, ensuring ethical data collection, verifying dataset quality, and assessing compliance with legal and regulatory requirements.
- Subject Matter Expertise (SME): The SME domain knowledge ensures that the data is relevant and of high quality, which is critical for training accurate and reliable AI models.

The critical role of data dependency in AI projects led to the creation of AI data management frameworks, such as Datumaro [16]. These frameworks provide tools for dataset versioning, format conversion, transformation, annotation management, and quality assessment, which are essential for managing data lifecycle in AI development.

*2) Recommendations*

We give the following tailoring guidance by performance domains to better address the Data Dependency aspect of AI project projects:

Table 1 Data Dependency Tailoring Guidance

| Domain | Recommendations |
|---|---|
| Stakeholder | Facilitate collaboration between data scientists, legal advisors, subject matter experts (SME), and end-users to align data deliverables with project objectives. Ensure transparent communication with stakeholders regarding data privacy, compliance, and ethical concerns. |
| Team | Specialized skills: To address the complexities of the projects and build multidisciplinary teams that include data scientists, legal experts, and SMEs. Education: Train project teams on the data lifecycle (sourcing, preprocessing, validation) and the specific data quality challenges, such as bias detection and mitigation. |
| Development Approach and Life Cycle | Introduce a structured framework for managing data-centric AI projects, defining stages for data sourcing, cleaning, preparation, and validation as part of the project lifecycle. See section 4.3 for more considerations on the frameworks. Embed iterative development practices to accommodate changes in data requirements, ensuring alignment with project objectives. |
| Planning | Plan for identifying and managing legal, ethical, and regulatory constraints, such as data ownership and licensing considerations or compliance with GDPR, CCPA, and similar regulations. |
| Project Work | Define processes for implementing robust data pipelines, including automated data quality checks and real-time validation mechanisms. Establish workflows for data preprocessing and alignment with project objectives to ensure efficient and effective project execution. |

| Domain | Recommendations |
|---|---|
| | Perform Privacy Impact Assessments (PIAs). Include processes for monitoring compliance with licensing agreements and intellectual property rights in delivered data products. |
| Delivery | Specify guidelines for data-driven solutions, ensuring datasets meet accuracy, completeness, and ethical standards. |
| Measurement | Integrate metrics and KPIs specifically for data quality, such as completeness and accuracy benchmarks and dataset bias and fairness measures. Provide tools and techniques to assess and track the reliability and alignment of data with project goals. Outline methods for statistical validation and automated quality checks to ensure data accuracy and completeness. |
| Uncertainty | Incorporate risk management strategies for addressing uncertainties in data projects, e.g., bias and ethical concerns in AI datasets, risks associated with legal compliance failures, or data breaches. Embed contingency plans to mitigate risks arising from errors in training datasets or system failures caused by poor data quality. |

## B. 4.2. Uncertainty and Experimentation (PF2)

### 1) Analysis

While the PMBOK Guide emphasizes the importance of managing uncertainty for project success and presents various management techniques such as uncertainty identification and assessment, risk management, adaptive planning, team capability enhancement, and stakeholder management, it falls short in providing specific guidelines for the unique challenges of AI projects, including the inherent uncertainty of model predictions, iterative trial and error, and the difficulty of securing sufficient resources [5] [29] [30].

- Unpredictable outcomes: The behavior of AI models can be complex and unpredictable. The earlier stages address the degree of uncertainty in traditional project management. Due to data issues and mathematical complexity, the uncertainty can remain constant or even increase as the experiments are performed [8].
- Experimentation: Experimentation with different algorithms, hyperparameters, and data sets is crucial for achieving optimal results. It leads to changes even during iterations, making it hard to complete tasks within one iteration and/or maintain consistent sprint duration [8] the case study of the OpenVINO Training Extensions [15] project, which implements cutting-edge AI technology, clearly shows this.
- Tolerance for failure: A willingness to embrace failure and learn from mistakes is essential for successful AI development.
- Complexity of issues: Debugging AI algorithms is significantly more complex than traditional software. It requires extensive efforts to identify root causes and sufficient time and resources for retraining, experimentation, and other related activities ("Black Box problem").
- Unclear target goal: The lack of direct competitors or precise market specifications can make it challenging to define the goals and scope of algorithm development, hindering the establishment of a precise development timeline.
- Unlike traditional software, AI performance can vary with changing conditions, making fixed testing criteria insufficient. Bias, distribution shifts, and adversarial risks further complicate validation, requiring continuous monitoring, rigorous testing, and human oversight to ensure reliability and fairness. There is a body of work for quality assessment of AI software, e.g. [37], and work for particular areas of AI, e.g., for Large Language Models (LLM) [38] [39].
- Rapidly evolving technology: rapid evolution, fast innovation, continuous experimentation with new models and algorithms, and the emergence of novel challenges that require adaptive solutions uniquely characterize AI projects. For example, the progress in Language Models halved the required compute approximately every 8 months [40].

### 2) Recommendations

Given the prevalence of "unknown unknowns" in AI projects, where business needs and AI models are subject to continual change, an iterative, experimental approach to AI software development is needed in the development process [41] [42]. The following tailoring guidance can supplemented to better address the unique aspects of AI project management [37] [42] [43] [22] [44]:

Table 2 Uncertainty & Experimentation Aspect Tailoring Guidance

| Domain | Recommendations |
|---|---|
| Stakeholder | Educate on AI projects: Regularly educate stakeholders on AI projects' experimental and uncertain nature to align expectations, foster adaptability, and communicate model limitations and risk mitigation strategies to address potential failures or errors. Promote Transparency: Share AI quality evaluation results with all relevant stakeholders. Inform stakeholders so that they can understand the project's progression and be aware of potential missed deadlines [22]. Foster collaboration between technical and non-technical stakeholders including subject matter experts (SME). |
| Team | Specialized skills: Establish a team comprising data scientists, ethicists, legal counsel, security experts, software engineers, and business analysts and ensure well-defined roles [43]. Enhance Interdisciplinary Communication: Facilitate regular interactions and establish a shared vocabulary among team members to bridge knowledge gaps [22]. Education: Facilitate continuous learning to stay current with rapidly evolving technologies. Ensure managers understand software development and machine learning and can align product and model teams [22]. |
| Development Approach and Life Cycle | MVP Approach: Employ an iterative experimentation process to continually enhance AI models and validate through MVPs [42]. |
| Planning | Options- and Risk-based planning: Develop flexible plans considering various scenarios to respond quickly to changes. Establish a risk-aware planning approach from the early stages of AI projects [43]. Keep extra buffer time and add additional timeboxes for R&D in the initial phases [22]. |
| Project Work | Continuous Validation: Automate the training-to-deployment AI pipeline using CI/CD for efficiency. Validate the model performance and accuracy of the final output using real-world data and cases to ensure business value [44] and consider A/B Testing [37]. |
| Delivery | Requirements: Adopt an incremental approach to exploring and refining AI requirements. Divide each requirement into multiple phases, with experiment results crucial to defining subsequent phases. Involve data scientists in the early stages of requirements solicitation [29]. Continuous Improvement: Collect user feedback, elicit requirements, and continuously improve the model [37] [42] [43]. |

| Domain | Recommendations |
|---|---|
| | AI Quality Standards and Transparency: Establish clear quality standards for AI-powered software deployments. Employ Explainable AI techniques to enhance the predictability and trustworthiness of AI model outputs [37].<br>Post-deployment refinement: Integrate automated experimentation and evaluation into the CI/CD pipeline to facilitate continuous improvement of AI models [42]. |
| Measurement | KPIs: Set Key Performance Indicators (KPIs) based on customer and subject matter experts (SME) feedback and real-world datasets and monitor them regularly [37] [42] [43].<br>Metrics: Establish AI quality evaluation criteria, continuously measure accuracy, performance, reliability, explainability, bias, and fairness, and align with business performance. Regularly conduct A/B testing and model validation [42] [43].<br>Quality Evaluation Automation: Implement automated evaluation tools and continuous monitoring systems within the CI/CD pipeline. Beware of the resource consumption for compute-heavy AI training jobs [37]. |
| Uncertainty | Inherent Limitations: Acknowledge AI models' inherent limitations and implement strategies to minimize risks during failures or errors [43].<br>Risk Management: Develop a risk assessment checklist to identify potential quality issues such as lack of generalization performance, model opacity (Black-Box Problem), security and safety issues, and legal concerns [37]. Identify and mitigate AI system uncertainty in data quality, model reliability, and security. Addresses AI uncertainty through automated testing frameworks and continuous CI/CD pipeline validation processes. |

## C. Iterative Development (PF3)

### 1) Analysis

Traditional software components like user interfaces and customer-facing features are particularly well-suited to Agile. These benefit from continuous refinement based on user feedback and evolving needs. Agile breaks development into smaller, predictable sprints, enabling teams to adapt to changing requirements and incorporate feedback quickly.

The AI components, however, require a more flexible, experiment-driven approach with unpredictable timelines. Estimating time accurately is often a struggle [22], as model tuning and experimentation can take longer than expected, making traditional sprint planning challenging [23]. This results in prompt mid-course adjustments within a sprint, while other experiments may span multiple sprints [8] [24]. The work [25] stresses that rigid interpretations of agile software development processes are poorly fit for AI projects.

The PMBOK Guide strongly emphasizes adaptability and tailoring project management practices to suit specific contexts. Despite its adaptability, the PMBOK Guide falls short in several areas that are critical for managing iterative development in AI projects:

- Lack of specific methodology guidance: While iterative approaches are acknowledged, the guide does not recommend or reference Agile or AI frameworks.
- Incomplete alignment with AI needs: It does not address unique challenges such as the continuous retraining of models based on evolving data, managing performance feedback loops, or mitigating model drift. Furthermore, it lacks detailed practices for frequent experimentation, a core requirement for AI development.
- Inadequate coverage of feedback mechanisms: Although iterative processes are encouraged, the guide does not emphasize mechanisms for incorporating continuous feedback into development cycles, a critical aspect of managing AI projects effectively.

### 2) Recommendations

Several tailoring recommendations strengthen the development approach and address the gaps in managing iterative development for AI software projects.

Balancing two traditional software developments with experiment-driven AI component research is necessary. The traditional software components benefit from comprehensive frameworks for Scrum, Kanban, and Agile approaches [9] [45] [46]. The AI research parts are less predictable and may span over a few agile sprints or rapidly change direction. The lifecycle shall adapt to the uncertain nature of the research. Project team can leverage a set of emerging process models [10] in AI research, including Cross-Industry Standard Process for Machine Learning (CRISP-ML(Q)) [47], Machine Learning Workflow (MLW) [48], Data Science Workflow (DSW) [49], Machine Learning Lifecycle (MLLC) [50], Process for Developing and Deploying AI models (PDDA) [51], MLOps [52] [28].

Establishing feedback loops and metrics that enable continuous refinement of AI models is also essential. The project team shall develop guidelines for setting up and managing these feedback loops, leveraging performance metrics such as accuracy, precision, and recall to monitor model improvements. Furthermore, mechanisms for rapid experimentation and evaluation of results should facilitate swift iterations and ensure that the project remains aligned with its objectives.

The table below provides recommendations for tailoring to address the specifics of the iterative development of AI projects.

Table 3 Iterative Development Tailoring Guidance

| Domain | Recommendation |
|---|---|
| Stakeholder | Regular Reviews: Engage stakeholders in sprint reviews to validate model performance and capture feedback. |
| Team | Team Collaboration: Define roles for iterative workflows within the AI project context. Highlight techniques for effective communication and collaboration in multidisciplinary teams. |
| Development Approach and Life Cycle | Hybrid lifecycle: Synchronize a balanced, agile approach for traditional software components with an experimental approach for AI components. Adopt a hybrid lifecycle that addresses both approaches. Encourage quick prototype development to validate ideas and identify potential issues early. Establish iterative testing and retraining cycles.<br>Minimum Viable Product (MVP) Approach: Recommend incremental delivery of AI components to enable early feedback and continuous improvement. |
| Planning | Adaptive Planning Techniques: Define processes for dynamic backlog updates based on model performance results and stakeholder feedback. |
| Project Work | Quality Assurance: Establish monitoring and experimentation infrastructure with automated pipelines for model training, testing, and deployment to streamline iterations. |
| Measurement | Key Performance Indicators (KPIs): Develop KPIs specific to iterative AI projects, such as time per iteration, model improvement metrics (e.g., reduction in error rates), cycle time for experimentation, and feedback loops. |

| Domain | Recommendation |
|---|---|
| | Tracking Progress: Use visual tools like burndown charts and cumulative flow diagrams to monitor iterative progress. |

### D. Specialized Expertise (PF4)

*1) Analysis*

AI software projects demand specialized expertise, particularly in fields like artificial intelligence (AI) and machine learning (ML):

- Multidisciplinary Teams: teams with diverse, hyper-specialized skills in AI and machine learning develop AI projects. These teams often consist of data scientists, software engineers, and subject matter experts (SME) whose expertise spans various stages, such as data preprocessing, algorithm design, and deployment. The authority hinges on technical or domain knowledge and academic expertise, not managerial leadership [4]. The PMBOK Guide does not provide clear guidelines on defining roles or coordinating cross-disciplinary efforts in AI/ML projects, where collaboration between individuals with specialized knowledge is critical for success. As AI projects are often complex and iterative, communication barriers frequently arise between solo contributors and technical and non-technical team members, exacerbating the difficulty of managing such diverse groups [31].
- Specialized Tools and Technologies: The AI projects require a range of advanced tools and platforms to manage the complexities of AI and machine learning workflows, which include AI/ML frameworks (e.g., TensorFlow, PyTorch, Scikit-learn), big data technologies (e.g., Hadoop, Apache Spark), and cloud computing platforms (e.g., AWS, Azure, Google Cloud Platform). These technologies are essential for model training, deployment, data processing, and maintaining large-scale AI infrastructure. Additionally, there is a need for tools for project management, data management, model performance monitoring, or compliance, all of which play a critical role in the success of AI projects. Without proper guidance on selecting, integrating, and maintaining these specialized tools, project teams may face inefficiencies, delays, and misalignment with the project's technical needs [4].
- Talent Acquisition and Skill Development: AI projects require specific expertise that may be in limited supply, making talent acquisition a significant challenge. The scientific knowledge is mainly gained in higher-level education (e.g., Doctoral studies). continuous upskilling is necessary to ensure project teams keep pace with the rapidly evolving AI/ML technologies. The fast-changing landscape of AI technologies demands that project team members stay updated on the latest advancements to remain competitive and effective.

*2) Recommendations*

The following table presents tailoring recommendations for specialized expertise by performance domains.

Table 4 Specialized Expertise Tailoring Guidance

| Domain | Recommendations |
|---|---|
| Team | Multidisciplinary Teams [4]: Structure multidisciplinary teams with clearly defined roles, such as data engineers, ML specialists, and subject matter experts (SME). Foster mutual understanding and exchange.<br><br>Talent Acquisition: Identify and recruit specialized skills for AI projects, including forming partnerships with universities or research laboratories, which can serve as valuable sources of talent and cutting-edge knowledge.<br><br>Skill Development: Build internal expertise through training programs, certifications, and knowledge-sharing sessions to ensure teams' readiness for the complexities of AI product development.<br><br>Team Communication: Facilitate cross-functional communication using shared vocabularies, visual aids, and Agile methodologies to bridge knowledge gaps. |
| Project Work | Select and implement specialized tools:<br><br>Project Management Software: Tools like Jira, Trello, or Asana, tailored to Agile methodologies, can improve AI project management by helping teams manage tasks, track progress, and facilitate collaboration within cross-functional teams. Agile methodologies are particularly suited for iterative AI development, allowing for rapid adjustments and collaboration among multidisciplinary teams.<br><br>AI Development Platforms: AI-specific platforms and frameworks, such as TensorFlow, PyTorch, and Azure Machine Learning, provide the necessary infrastructure for building and deploying AI models.<br><br>Data Management Tools: One of the primary challenges in AI projects is managing large volumes of data. Data collection, preprocessing, and storage tools like Apache Hadoop, Databricks, or Google BigQuery are essential for data quality and scalability.<br><br>Monitoring and Analytics Tools: Tools such as Prometheus, Grafana, and MLflow provide essential capabilities for monitoring model performance and operational metrics. Real-time monitoring and visualization allow teams to identify issues, track changes, and optimize model performance throughout deployment.<br><br>Ethics and Compliance Tools: Ethical considerations and regulatory compliance are increasingly important in AI projects. Tools that assist in managing data privacy, detecting algorithmic bias, and documenting compliance efforts are crucial (see Section 4.5). |

### E. Ethical Considerations (PF5)

*1) Analysis*

AI is revolutionizing industries, offering immense potential and raising significant ethical concerns. Its impact goes beyond technology, influencing individuals, societies, and the environment. Recent controversies, such as Clearview AI's unauthorized facial recognition database [53], which led to legal actions and fines for privacy violations, and Amazon's biased AI hiring tool [54], which demonstrated gender discrimination, highlight the risks of unregulated AI development. The exponential growth of the AI incidents and growing body of AI related regulation [55] underscore the need for a holistic approach to ethical AI, considering the intricate interplay between humans and AI systems [29] [32] [33] [34] [35]. Based on the literature review and the case studies, we see the following key ethical considerations for AI projects:

- Bias and fairness: Bias in AI refers to systematic algorithm errors that lead to unfair or discriminatory

outcomes for certain groups, often based on factors like race, gender, or socioeconomic status. Fairness in AI aims to mitigate these biases and ensure that AI systems treat all individuals equitably.
- Transparency and explainability: Transparency in AI makes the underlying algorithms and decision-making processes understandable. Explainability provides clear and understandable explanations of how an AI system arrives at a particular decision, enhancing trust and accountability.
- Privacy and data protection: Privacy and data protection in AI involve safeguarding sensitive user data from unauthorized access, use, or disclosure. It is implemented by robust security measures, obtaining informed consent, and ensuring compliance with relevant data privacy regulations (e.g., GDPR, CCPA).
- Safety and robustness: Safety and robustness in AI refer to the ability of AI systems to operate reliably and predictably without causing harm. It is implemented by ensuring AI models' accuracy, reliability, and resilience and preventing unintended consequences or malfunctions.
- Human oversight and control: Human oversight and control in AI involve maintaining appropriate levels of human supervision and intervention in AI systems. They ensure AI systems' responsible and ethical use and that humans retain ultimate control over their decision-making.
- Accountability: Accountability in AI refers to the ability to identify and assign responsibility for AI systems' actions and outcomes. It is important to establish clear lines of accountability for AI developers, deployers, and users and mechanisms for addressing potential harms caused by AI.
- Sustainability: The use of AI involves the consumption of energy and computational resources, which can have environmental impacts. The project team shall consider These impacts when using AI in projects or specific tasks. Factors to consider are the AI models' energy efficiency, the data centers' carbon footprint, and the potential for AI to contribute to sustainable solutions, such as optimizing energy consumption or reducing waste.
- Licensing: understanding the licensing rights associated with the data used to train the AI model is crucial to AI-generated content. This includes data used for input and the resulting output. While copyright considerations are relevant, the primary concern often revolves around the licenses granted for using this data. Determining whether the necessary licenses are in place for training data and AI output is essential to avoid legal complications.

*2) Recommendations*

Below, we provide tailoring recommendations for the PMBOK Guide to better address the unique ethical aspects of AI projects:

Table 5 Ethical Aspects Tailoring Guidance

| Domain | Recommendations |
|---|---|
| Stakeholder | Ethical stakeholder engagement: Involve ethicists, regulators, advocacy groups, and the public to address ethical considerations throughout the project lifecycle. |
| | Promote transparency: Communicate AI components, potential impacts, and ethical considerations to stakeholders, ensuring informed participation. Feedback loops: Establish mechanisms for affected communities to provide input and monitor project outcomes, mitigating adverse effects over time. |
| Team | Specialized skills: Recognize and secure expertise in AI ethics to augment the hyperspecialized team with data science, machine learning, and business analysis expertise. |
| Project Work | Ethical principles integration: Embed fairness, transparency, and accountability into all project lifecycle phases. |
| Delivery | Monitoring systems: Implement systems for tracking AI system performance and ethical compliance after deployment. |
| Measurement | Ethical impact assessments: Regularly assess the project for ethical risks and societal impacts. Fairness audits: Establish measurable criteria for fairness and implement routine audits to track performance. Bias audits: Conduct bias audits regularly to ensure fairness and accuracy in AI outputs. Social impact assessments: Evaluate societal risks (e.g., algorithmic discrimination, job displacement) throughout the project lifecycle, adjusting objectives to mitigate these issues and conducting post-delivery evaluations to measure impacts. Continuous monitoring: Establish processes for ongoing evaluation of ethical and technical performance, ensuring long-term alignment with ethical principles. |
| Uncertainty | Ethics risks: Identify risks such as data bias, model drift, and algorithmic errors (see Section 4.2 for more areas to consider). Risk mitigation: Develop mitigation strategies for AI-related risks, such as fairness audits, ethical guidelines, and robust testing procedures. |

V. CONCLUSIONS

The rapid evolution and adoption of Artificial Intelligence (AI) technologies present significant challenges for traditional project management frameworks, such as the PMBOK Guide. While the PMBOK Guide provides a robust foundation for managing projects, its principles often fail to address the unique complexities of AI projects. These include data dependency, iterative and experimental development processes, specialized expertise requirements, and critical ethical considerations.

This paper has highlighted the misalignments between the PMBOK Guide and the demands of AI projects, offering detailed analyses and recommendations to bridge these gaps. By incorporating data-centric practices, embracing iterative methodologies, fostering interdisciplinary collaboration, and embedding ethical principles into the project lifecycle, the PMBOK Guide can better meet the needs of AI-driven initiatives.

Ultimately, as AI reshapes industries, project management frameworks must evolve alongside it. Integrating flexible, adaptive, and ethical approaches will improve the management of AI projects and ensure their alignment with societal values and the sustainable development of technology. These efforts will enable the PMBOK Guide to remain relevant and effective in a future dominated by intelligent systems.

Future research should explore the PMBOK Guide 8th Edition, which is undergoing revision at publication, to assess its adaptability to AI projects. Additionally, examining emerging AI project management frameworks, evolving ethical guidelines, and new AI regulations will help refine and update the recommendations for tailoring the PMBOK Guide to the unique demands of AI software projects.